\documentstyle{article}
\begin{document}
\title{Dimensional reduction in supersymmetric field theories}
\author{Oleg V. Zaboronsky \thanks{On leave from the Institute
of Theoretical and Experimental Physics, Moscow, Russia. } \\
Department of Mathematics, University of California at Davis, \\
Davis, CA 95616; e-mail: zaboron@math.ucdavis.edu}
\maketitle

\newcommand{\al}{\alpha}
\newcommand{\be}{\beta}
\newcommand{\rqu}{{\cal R} _{Q}}
\newcommand{\fti}{\tilde{\Phi}}
\newcommand{\Fti}{\tilde{\Phi} (\Phi _{0} )}
\newcommand{\f}{\Phi}
\newcommand{\th}{\theta}
\newcommand{\tb}{\overline{\theta}}
\newcommand{\s}{\Sigma}
\newcommand{\sn}{\Sigma _{0}}
\newcommand{\ab}{\overline{a}}
\newcommand{\cb}{\overline{c}}
\newcommand{\Kb}{\overline{K}}
\newcommand{\da}{\frac{\partial}{\partial y^{\alpha}}}

\begin{abstract}
A class of quantum field theories
invariant with respect to the action of an odd vector field $Q$
on a source supermanifold $\Sigma$ is considered. We suppose
that $Q$ satisfies the conditions of one of localization theorems
(see e. g. \cite{SZ}). The $Q$-invariant sector of a field
theory from the class above is
 shown then to be equivalent to the quantum field
theory defined on the zero locus of the vector field $Q$. 
\end{abstract}
\section{Introduction}
The aim of the present paper is to connect the 
phenomenon of dimensional
reduction in supersymmetric field theories with 
localization of certain integrals
over supermanifolds. Let us start with the 
explanation of the meaning we assign to the
terms "dimensional reduction" and "localization"
(each of  these terms
is used in the literature in quite a 
few different contexts). 
By dimensional reduction we will 
understand a fact of an 
{\em exact} equivalence between  
a quantum field theory and another quantum field theory defined
on the submanifold of  the source manifold of the original theory. 
An example of such phenomenon is provided by the
celebrated Parisi-Sourlas model \cite{PS} which also served as a
main motivation for the present work. 

Localization of an integral over a (super)manifold
 $\Sigma$ to a subset $R \subset \Sigma$
means more or less that this integral is independent
of the values of the integrand on the complement
to the arbitrary neighborhood of $R$ in $\Sigma$. 
In what follows we will be
using the notion of localization in the even
more restricted sense. It is well-known
that localization is usually related
to the presence of some odd symmetry
of the problem. 
So
let $Q$ be an odd vector field
 on $\Sigma$. This means that $Q$ is a parity-reversing
derivation on $Z_{2}$-graded algebra of functions on $\Sigma$. 
We say that $Q$ satisfies the conditions of some localization
theorem if for any $Q$-invariant function $f$ on $\Sigma$
\begin{eqnarray}
\int_{\Sigma} dV \cdot f =\int_{R_{Q}} dv_{Q} f \mid _{R_{Q}}, 
\end{eqnarray}
where $dV$ is a fixed volume element on $\Sigma$; zero locus
of $Q$  is supposed to be a submanifold of
${\Sigma}$ and is denoted
by $R_{Q}; ~dv_{Q}$ stands for the volume element on
$R_{Q}$ depending on $dV, ~ Q$, but not $f$. 

An exposition of different localization techniques in the context of 
quantum field theory can be found in \cite{RS}. In \cite{SZ} we studied
localization in the framework of supergeometry. We managed to prove 
a general
localization theorem which includes essentially all
 previously known localization statements as its
 particular cases. The main result of \cite{SZ} can be formulated as
follows. Let $Q$ be an odd vector field on $\Sigma$
 which preserves a volume element
$dV$ on $\Sigma$ . Suppose that
 $Q^2 =\frac{1}{2} \{ Q, Q \}$ belongs to a Lie
algebra of a compact subgroup of the group of
 diffeomorphisms of $\Sigma$. Then under
some additional conditions of non-degeneracy
 of $Q$ the integrals of $Q$-invariant functions
 over $\Sigma$ localize to the zero locus $R_{Q}$
 of the vector field $Q$. 
In other words (1) holds with $dv_{Q}$ determined
by $dV$ and the matrix of the first derivatives of the vector
field $Q$ at $R_{Q}$.

To conclude the Introduction let us formulate
 and prove another localization theorem 
which will be useful in the analysis of
 dimensional reduction of Parisi-Sourlas-type models. 

The explanation of basic notions of supergeometry
which will be used below can be found in \cite{S2}. 

{\bf Theorem. } {\em Let $\Sigma$ be a compact
supermanifold equipped with an even metric
 $g$. Suppose $Q$ is
an odd vector field on $\Sigma$ preserving
 the metric, i. e. $L_{Q} g =0$. 
Suppose that vector field
$Q^2$ is non-degenerate in the
 vicinity of its zero locus
$R_{Q^2}$. Suppose also
 that odd and even codimensions
of  $R_{Q^2}$ in $\Sigma$ coincide. Then for
 any $Q$-invariant function $f$ on $\Sigma$
\begin{eqnarray}
\int_{\Sigma} dV f = \int_{R_{Q^2}} dv_{Q} 
f \mid _{R_{Q^2}}, 
\end{eqnarray} 

where $dV$ is a volume element on
 $\Sigma$ corresponding to the metric $g$ and
 $dv_{Q}$ is a volume element
on $R_{Q^2}$ determined completely
by $g$ and $Q$. } \footnote{Initially this theorem was
 formulated and proved
for the linear superspaces. Its present form
 benefits from the collaboration with A. S. Schwarz}. 

{\em Proof. }Let $\{ z ^{\al} \} $
 be a set
 of local coordinates on $\Sigma$. The parity of the
$\al$'th coordinate will be denoted by $\epsilon _{\al}$. 
In these coordinates 
$Q=Q^{\al} (z) \frac{\partial}{\partial z^{\al}}, 
Q^2=(Q^2) ^{\al} (z) \frac{\partial}{\partial z^{\al}}$, 
where $(Q^2) ^{\al} =Q ( Q^{\al} (z))$.
 We will write the metric in the form
$g=g_{\al ~ \be} (z) \delta z^{\al} \delta z^{\be}$.
Consider now an odd
function $\sigma$ on $\Sigma$ defined in the local 
coordinates by the following expression: 
\begin{eqnarray}
\sigma (z) = \frac{1}{2} \sum _{\al , \be } (-1) ^{\epsilon _{\al}
+\epsilon _{\be}} g_{\al ~ \be} Q^{\al} (z) (Q^2) ^{\be} (z)
\end{eqnarray}
  
It is easy to verify that the right hand side of (3) does not
depend on the choice of  local coordinates, so that $\sigma$
is indeed a function on $\Sigma$. A direct calculation
shows that $\sigma$ is $Q^2$-invariant, i. e. $Q^2 \sigma =0$. 
Also one finds that $Q \sigma (z) = g_{\al ~ \be} (Q^2) ^{\al}(z)
 (Q^2) ^{\be} (z) \equiv < Q^2 (z), Q^2 (z) >$, 
where $<, >$ denotes the pairing in the fibers of
the tangent bundle over $\Sigma$ induced by the
metric $g$. 

Another computation shows that $R_{Q^{2}}$ is a
subset of the critical set of $Q \sigma$, i. e. 
$\nabla Q \sigma \mid _{R_{Q^{2}}}=0$. It follows
from non-degeneracy of $Q^2$ in the
vicinity of $R_{Q^2}$ that $R_{Q^2}$
is a non-degenerate critical set. The last means that
the Hessian of $Q \sigma$ has the maximal rank at each
point of  $R_{Q^2}$. 

Our aim is to compute $\int_{\Sigma} dV f$, where $Q f =0$. 
The fact that the metric $g$ is $Q$-invariant implies that
$div_{dV} Q =0$, which means that the volume
element on $\Sigma$ constructed using the $Q$-invariant
metric is also $Q$-invariant. 
Thus it is easy to see that the following
is true: 
 \begin{eqnarray}
\frac{\partial}{\partial \lambda}
\int_{\Sigma} dV \cdot f e ^{- \lambda Q \sigma} = 
- \int_{\Sigma} dV  Q ( f e^{- \lambda Q \sigma} ) =0
\end{eqnarray}
Let $\{ U_{m} \} _{m \in I }$ be a finite atlas of $\Sigma$, 
$\{ h_{m} \} _{m \in I}$ - a partition
 of unity on $\Sigma$ subordinate
to this atlas. Suppose also that the atlas is chosen to satisfy
the following two condition: 

(i) if  $\overline{U_{k}} \bigcap
R_{Q^2} \neq \emptyset , k\in I, $ then $U_{k}
\bigcap R_{Q^2} \neq \emptyset$; 

(ii) if $U_{k} \bigcap R_{Q^2} \neq \emptyset$
then the critical set of $Q \sigma \mid _{U_{k}}$
is just $U_{k} \bigcap R_{Q^2}$. 

Using (4) one can rewrite an expression for
the integral of $f$ over $\Sigma$ as follows: 
\begin{eqnarray*}
 \int_{\Sigma} dV ~ f = \lim_{\lambda \rightarrow \infty}
\int_{\Sigma} dV ~ f e^{- \lambda <Q^2, Q^2>}=   \\
\sum_{m \in I} \lim_{\lambda \rightarrow \infty} 
\int_{U_{m}} dV h_{m} f e^{- \lambda <Q^2, Q^2 > }=   
\end{eqnarray*}

Let $m(Q^2)$ denotes a number part of the vector field
$Q^2$ , 
$R_{m(Q^2)} \subset \Sigma$ denotes zero locus of $m(Q^2)$. 
Let us choose $k \in I ~ : ~ \overline{U_{k}}
 \bigcap R_{Q^2} = \emptyset$. 
Then $\overline{U_{k}} \bigcap R_{m(Q^2)} = \emptyset$. 
As a consequence of (i)
the number part of $<Q^2, Q^2>$ is positive at each point
of $U_{k}$, so one can find such positive constants $c_{1}$
and $c_{2}$ that: 
\begin{eqnarray*}
\mid \int_{U_{k}} dV h_{k} f e^{- \lambda <Q^2, Q^2> } \mid
\leq c_{1} \lambda ^{n} e^{- c_{2} \lambda } \rightarrow
0~ \mbox{ as } ~ \lambda \rightarrow \infty
\end{eqnarray*}
Thus we conclude that
\begin{eqnarray}
\int_{\Sigma} dV ~f = \sum_{ \{ k \in I \mid U_{k} \bigcap R_{Q^2}
\neq \emptyset \} } \lim_{\lambda \rightarrow \infty }
\int_{U_{k}}dV h_{k} f e^{- \lambda <Q^2, Q^2>}
\end{eqnarray}
The integrals in the right hand side of  (5) can be calculated using
the Laplace method adapted to include integrals
over superspaces (see e. g. \cite{S1} ). Under
the condition that odd codimension of $R_{Q^2 }\subset \Sigma$
is equal to its even codimension we obtain that
\begin{eqnarray}
\lim_{\lambda \rightarrow \infty} \int_{U_{k}} dV h_{k} f
e^{-\lambda < Q^2 , Q^2 >} = \int_{U_{k} \bigcap R_{Q^2}}
dv_{Q} (h_{k} f) \mid_{U_{k} \bigcap R_{Q^2}}, 
\end{eqnarray}
where $dv_{Q}$ is the volume element on $U_{k} \bigcap R_{Q^2}$
defined as a partition function of degenerate
 functional $ Q \sigma \mid_{U_{k}}$
(see \cite{S1}, lemma 2). By (5) $Q \sigma$ depends on $Q$
and $g$ only, so does $dv_{Q}$. It remains to
notice that the set $\{ h_{k} \mid  _{R_{Q^2}} \} _{ \{ k \in I
\mid U_{k} \bigcap R_{Q^2} \neq \emptyset \} }$ provides one
with the partition of unity on $R_{Q^2}$. Therefore substituting
(6) into (5) and using the definition 
of the integral over a (super)manifold
we see that
\begin{eqnarray*}
\int_{\Sigma} dV \cdot f = \int_{R_{Q^2}} 
dv_{Q} \cdot f \mid _{R_{Q^2}}
\end{eqnarray*}
The Theorem is proved. 

{\bf Corollary. } {\em If  $R_{Q^2}=R_{Q}$ the Theorem
implies the localization of corresponding integrals
over $\Sigma$ in the sense of the definition (1). }

Note that the statement of the 
Theorem above can be formally justified
in the case when $\Sigma$ is an infinite-dimensional
manifold. For example $\Sigma$ can be realized as a space
of maps from a world sheet to a target
manifold of some quantum field theory. This suggests
that there are possible applications of the  Theorem
above which are different from the ones we consider below.   

Finally let us remark that if $R_{Q^2} \neq R_{Q}$ 
one can still prove the localization
of the integrals under consideration
to $R_{Q}$. The proof will consist of two steps: 
first one repeats the arguments
above to prove the localization to $R_{Q^2}$. 
Then one notes that the vector field $Q$ generates a nilpotent
vector field on $R_{Q^2}$ and $f \mid _{R_{Q^2}}$ is
invariant with respect to this vector field. Corresponding
integral is localized to $R_{Q}$ ( see e. g. \cite{Wit}, 
\cite{S4} ). 

\section{Derivation of the main result. }

Let $\Sigma$ be a compact supermanifold. 
Suppose that $Q$ is an
odd vector field on $\Sigma$. 
We always assume that zero locus $R_{Q}$ of the vector
field $Q$ is a submanifold of $\Sigma$ and that
$Q$ is non-degenerate in the neighborhood of $R_{Q}$. 
Let $dV$ be a fixed $Q$-invariant
volume element on $\Sigma$. 
Assume that $Q$ satisfies
the localization conditions, i. e. (1) holds for any $Q$-invariant
function $f$ on $\Sigma$. 
Let $M$ be another
supermanifold. To avoid irrelevant technicalities we suppose that
$M$ is diffeomorphic to a linear superspace. 
Denote by $E$ the (super)space of maps from $\Sigma$ to $M$. 
Naturally an action of $Q$ on $\Sigma$ generates
an infinitesimal diffeomorphism of the space of maps: 
\begin{eqnarray}
\Phi \rightarrow \Phi + \epsilon Q \Phi ~, 
\end{eqnarray}
where $\Phi \in E$ and $\epsilon$
is an odd parameter. 
We will use the notation $\hat{Q}$ for the vector
field on $E$ corresponding to (7). 

Next let us impose an additional condition
on $Q$ which will be crucial for further
considerations. Namely we will assume that the following
Cauchy problem has a solution: 
\begin{eqnarray}
Q \Phi =0 \\
\Phi _{R_{Q}} =\Phi_{0}, 
\end{eqnarray}
where $\Phi _{0}$ is any map
from $R_{Q}$ to $M$. In other words
we require that any map $R_{Q} \rightarrow
M$ can be continued to the
$Q$-invariant map $\Sigma \rightarrow M$. 
In all interesting cases the problem (8), (9)
has a lot of solutions. We suppose that
the space of solutions of (8) corresponding
to a fixed initial condition (9) is contractible. 

Consider now a quantum field theory defined
on $\Sigma$. Let
 ${\cal L} ( \Phi, \partial \Phi ), \Phi \in E$ be
a corresponding quantum Lagrangian. The 
word "quantum"  means
that having started from classical field theory
we fixed gauge-like symmetries 
of the classical Lagrangian
using some quantization procedure (BV for example, see \cite{BV})
and arrived at the expression for 
${\cal L} (\Phi , \partial \Phi)$ where $\Phi$ is a map
from $\Sigma$ to the manifold $M$ of both physical
and auxiliary fields such as ghosts, antifields, etc. 
Therefore corresponding action functional
is non-degenerate, i. e. the linear integral
operator
in $T_{\Phi} (E)$
 with the kernel $\frac{\delta ^{2} S}
{\delta \Phi ^{i} (x) \Phi ^{j}(y)}$ has
no zero eigenvectors for any $\Phi \in E$. 
Here $x~, y \in \Sigma$
and a choice of local coordinates in $M$ is assumed. 
 
The main condition imposed on the quantum field
theory at hand is $\hat{Q}$-invariance. 
Namely we suppose
that 
\begin{eqnarray}
{\cal L} (\Phi + \epsilon Q \Phi, 
\partial (  \Phi + \epsilon Q \Phi ))=
{\cal L} ( \Phi, \partial \Phi) + 
\epsilon Q {\cal L} ( \Phi , \partial \Phi )
\end{eqnarray}
The fact that the action $S=\int_{\Sigma} dV {\cal L}$
is $\hat{Q}$ is invariant
follows then from (10) and the $Q$-invariance of the volume
element $dV$. 

There is a simple construction generating 
a lot of models satisfying (10). 
Let $h_{n}$be a multivector field of rank $n$ 
on $\Sigma$. 
Introducing local coordinates $\{ z ^{\al} \} $
on $\Sigma$ one can present $h_{n}$ in the following form: 
\begin{eqnarray*}
h_{n}= h_{n}
^{\al _{1}
 ~ \al _{2} \ldots \al _{n}} (z) 
\frac{\partial}{\partial z^{\al _{1}}}
\otimes \frac{\partial}{\partial z^{\al _{2}}} \otimes \ldots
\otimes \frac{\partial}{\partial z^{\al _{n}}}
\end{eqnarray*}
Using a multivector
field $h_{n}$
and a map $\Phi : \Sigma
\rightarrow M$ one can construct a map $h_{* ~ n} $
from $\Sigma$ to the
$n'$th tensor power of the tangent bundle $TM$
over $M$. Choosing local coordinates both in $\Sigma$
and $M$ one can present it as follows: 
\begin{eqnarray}
h_{* ~ n} (z)=
 h_{n} ^
{\al _{1}
 ~ \al _{2} \ldots \al _{n}} (p) 
\frac{\partial \Phi ^{i_{1}}}{\partial z^{\al _{1}}}
\frac{\partial \Phi ^{i_{2}}}{\partial z^{\al _{2}}} \cdots
\frac{\partial \Phi ^{i_{n}}}{\partial z^{\al _{2}}} \nonumber \\
\equiv h_{n} ( \Phi ^{i_{1}} \times \Phi ^{i_{2}} \times
 \ldots \times \Phi ^{i_{n}} ) (z)
\end{eqnarray}

Suppose now that $h_{n}$ is $Q$-invariant, 
i. e. $L_{Q} h_{n}=0$, where $L_{Q}$ is
a Lie derivative with respect to the vector
field $Q$. Then as it is easy to see 
\begin{eqnarray}
\epsilon Q (h_{n} ( \Phi \times \ldots \times \Phi))= \\
= h_{n} ( (\Phi + \epsilon Q \Phi ) \times \ldots \times
(\Phi + \epsilon Q \Phi )) - h_{n} ( \Phi \times \ldots
\times \Phi)  \nonumber
\end{eqnarray}
 
Suppose finally that the derivatives of $\Phi$ enter
the Lagrangian only in the form of combinations (11), 
where   $h_{n}$ is a $Q$-invariant multivector field. 
Then in virtue of (12) the relation
(10) is satisfied and corresponding
action functional is $\hat{Q}$-invariant. 

Let us illustrate the above considerations with the
following example. Take $g$ to be a $Q$-invariant
metric on $\Sigma$. Then the following model
is $\hat{Q}$-invariant: 
\begin{eqnarray}
S=\int_{\Sigma} dV ( g^{ \al   \be } \partial _{\al} \Phi ^{i} 
 \partial _{\be} \Phi ^{j} G_{i   j} (\Phi) + V(\Phi ))
\end{eqnarray}
Here $g^{\al   \be}$ is a $Q$-invariant multivector field
of rank 2 inverse to the metric
tensor $g_{\al   \be}$ and $G_{i   j}$
is a metric tensor on $M$. Note that (10) constitutes a natural
non-linear generalization of Parisi-Sourlas model \cite{PS}. 

Now we are able to formulate the main result of the paper. 
The $Q$-invariant (Schwinger) 
correlation functions of the theory described above
have the following generating functional: 
\begin{eqnarray}
Z[J]=\int [D \Phi ] _{E} e^{i \be (S [\Phi] + \int_{\Sigma} dV
J_{i}(p)  \Phi ^{i} (p))   }, 
\end{eqnarray}
where
$\{ J_{i} \}$ are $Q$-invariant functions on $\Sigma$ playing a role
of sources, $[ D \Phi] _{E}$ is a formal measure 
on the space of maps $E$, 
$\beta$ is a coupling constant. 

By means of
formal manipulations
with functional integrals
we are going to show that under the conditions on $Q$
and $S[\Phi ]$
the generating functional (14) can be rewritten as follows: 
\begin{eqnarray}
Z [J]=\int [D \Phi] _{E_{Q}} exp^{i \be (S[ \Phi \mid _{R_{Q}}]
+\int_{R_{Q}} dv_{Q} J_{i} (p)  \Phi ^{i} (p) \mid _{R_{Q}})}
\end{eqnarray}
Here $E_{Q}$ denotes the space of maps from $R_{Q}$ to
$M$, $[D \Phi] _{E_{Q}}$ is a measure on $E_{Q}$; the new
action functional is 
\begin{eqnarray}
S [\Phi  \mid _{R_{Q}} ]
=\int_{R_{Q}} dv_{Q} {\cal L} (\Phi \mid _{R_{Q}}, 
\partial ^{\prime} \Phi
\mid _{R_{Q}}, 0)
\end{eqnarray}
where the new Lagrangian is obtained from the old one
by restricting the fields to $R_{Q}$ and setting the derivatives
of the fields in the directions transversal to $R_{Q}$ equal to $0$. 
We also used the symbol $\partial  ^{\prime}$ to denote the
derivatives along $R_{Q}$. 

Eq. (15) states the equivalence between the $Q$-invariant sector
of the initial theory and the theory determined by the action
functional (16) defined on the submanifold of the initial
source manifold $\Sigma$. According to the adopted terminology, 
dimensional reduction occurs. 

Note that in the case when $R_{Q}$ is 
zero-dimensional, the r. h. s
of (15) reduces to a finite-dimensional integral,
 which means an
exact solvability of the $Q$-invariant sector
 of the theory we have
started with. We also see that in the situation
 when $Q$ happens to have
no zeros at all the $Q$-invariant sector
 is trivial which yields a set
of Ward identities for the correlation
 functions of the initial theory.

To demonstrate the equality between (14) and (15) let us
consider first the subset ${\cal R} _{Q}$
of $E$ consisting
of $Q$-invariant maps from $\Sigma$ to $M$. The space $\rqu$
is foliated by means of the following equivalence relation: 
two $Q$-invariant maps $\Phi , \Phi ' \in \rqu$ belong
to the same fibre of the foliation iff $ \Phi \mid _{R_{Q}}=
\Phi ' \mid _{R_{Q}}$; in other words $\Phi$ and $\Phi '$
are equivalent if they determine the same element of $E_{Q} =
\{ R_{Q} \rightarrow M \}$. Consider a
section of such foliation - a map $\fti : E_{Q} \rightarrow
\rqu$ which assigns to each element of $E_{Q}$ {\em a unique}
element of $\rqu$. In other words we set a rule which
singles out one and
only one solution to the problem (8), (9) for
each $\Phi _{0}=\Phi _{R_{Q}}$. Such section exists
due to the stated assumptions about the
space of solutions of the problem (8), (9). 
Consider now the following functional
on $E$: 
\begin{eqnarray}
F [ \Phi ] =\int_{\Sigma} dV G_{ij} (\Phi ) 
( \Phi ^{i} - \Fti ^{i}) (\Phi ^{j} - \Fti ^{j} ), 
\end{eqnarray}
Clearly , $\hat{Q} F [\Phi ] =0$. 
Using (17) we introduce
the following deformation of the generating functional (13): 
\begin{eqnarray}
Z_{\lambda} [ J] = \int [ D \Phi ] _{E} e^{i \beta (S [\Phi]
+ \int_{\Sigma} dV  J_{i}  \Phi ^{i}
+\lambda F [ \Phi ]) }
\end{eqnarray}
Note that $Z [J] = Z_{0} [J]$. Let us show that $Z_{\lambda} [J]$
is in fact independent of $\lambda$: 
\begin{eqnarray}
\frac{\partial}{\partial \lambda} ln ~ Z_{\lambda} [J]=
i \beta \int_{\Sigma} dV < F [\Phi (p) ] >_{\lambda , J}, 
\end{eqnarray}
where $< ~ > _{\lambda , J}$ denotes the average with respect
to the "action" functional - 
an argument of exponent in (18). A correlator
$< F [\Phi (p)] > _{\lambda , J}$ can be considered as a function
on $\Sigma$. It follows from (17) that the restriction of this function
to $R_{Q}$ is zero. Moreover this function is $Q$-invariant
as a consequence of the $Q$-symmetry of the problem. Really, 
$Q < F[\Phi ] > _{\lambda , J } = 
< \hat{Q} F [\Phi] > _{\lambda, J}=0$. 
The last equality can be regarded as a Ward identity corresponding to
 the $Q$-invariance
of  the vacuum of the theory at hand.
 Thus the r. h. s of (19) is
an integral over $\Sigma$ of a $Q$-invariant function equal to zero
on zero locus $R_{Q}$ of $Q$.
 Therefore it is equal to $0$ in virtue
of localization condition (1). 

So, $Z_{\lambda} [J]$ is independent of $\lambda$. Thus
one can compute the generating function $Z [J]$ as follows: 
\begin{eqnarray}
Z [J] = \lim_{\lambda \rightarrow  \infty} Z_{\lambda} [J]
\end{eqnarray}
 One can rewrite the r. h. s of
(18) in the following form: 
\begin{eqnarray}
Z [J] = \lim_{\lambda \rightarrow
 \infty} \int [D \Phi _{0} ] _{E_{Q}}  
 \int_{ \{  \Phi _{R_{Q}} =\Phi _{0} \} } 
 [D \Phi ] _{E} e^{i \be (S[\Phi]  
+\int_{\Sigma} dV J_{i} \Phi ^{i} +\lambda
F[\Phi])}
\end{eqnarray}
In the limit  $\lambda \rightarrow \infty
$ the internal integral in (21) localizes to the 
the critical points of the functional
$S [\Phi ] + \int_{\Sigma} dV 
+\int_{\Sigma} dV  J_{i}\Phi^{i}+ \lambda F [\Phi]$
which is defined on the space of maps having a fixed
restriction to $R_{Q}$. It follows from
the $Q$-invariance of this functional that 
one of these critical points if $\Phi = \Fti$
( see \cite{S3} for a proof in the even case). 
It can be shown under very general
assumptions on $S[ \Phi ]$ that $\Phi = \Fti$
is the only extremum contributing
to (21) in the limit $\lambda \rightarrow \infty$. 
The contribution
can be calculated 
using infinite-dimensional version of 
the stationary phase method. As a result we obtain the
following answer for the generating functional  (14): 
\begin{eqnarray}
Z[J] = \int [D \Phi _{0}] _{E_{Q}}
e^{-\be S[\Fti ] + \int_{\Sigma} dV J_{i } \Fti ^{i} }, 
\end{eqnarray}
where we absorbed the determinants 
which appeared as a result of computation of corresponding
gaussian integrals into the redefinition 
of functional measure on $E_{Q}$. But now we
note that in virtue of (10) 
\begin{eqnarray*}
Q {\cal L} (\Fti , \partial \Fti) =0, 
\end{eqnarray*}
therefore the integral $S[\Fti ] 
=\int_{\Sigma} {\cal L} (\Fti , \partial \Fti)$ localizes
to the zero locus of the vector field $Q$. 
It follows also from the non-degeneracy of
$Q$ in the vicinity of $R_{Q}$ that 
$\partial _{\perp} \Fti \mid _{R_{Q}} =0$. 
This remark together with localization condition 
(1) permits us to conclude that
\begin{eqnarray}
S[\Fti ] = \int_{R_{Q}} dv_{Q} 
{\cal L} ( \Phi _{0} , \partial  ^{\prime } \Phi _{0}
, 0)
\end{eqnarray}
The same localization arguments work for the source 
term as we have chosen the functions
$J$'s to be $Q$-invariant. Substituting 
(23) into (21) we arrive at the expression (15)
for the generating functional of the reduced theory. 

The way we established the equality between 
(14) and (15) is somewhat
naive in the sense that the result was achieved
 by means of formal manipulations
with the path integral without addressing the questions of 
proper renormalization of the loop expansion arriving. 
Our results only suggest the possibility
of the phenomenon considered, an additional
analysis is required in each particular case.  
  
Keeping up with the level of generality adopted for the 
present section 
we can discuss the relation between instanton sectors
in the original and the reduced theory. Suppose that
$\f _{0}$ is an extremum of the action functional
$S_{red} [\f]$ of the reduced theory, $\Phi : R_{Q}
\rightarrow M$. Let $\Fti \in \{ \Sigma \rightarrow M  \}$
be a $Q$-invariant map such that its restriction to
$R_{Q}\subset \Sigma$ coincides with $\f _{0}$. 
Then $\fti$ is an extremum of the action
functional $S [\f]$ of the original theory. The proof
of this statement is based on the $\hat{Q}$-symmetry
of $S [ \f ]$ and goes along the same line as its
even counterpart (see \cite{S3}). Conversely, 
any $Q$-invariant extremum of the original theory
produces a solution to the equations of motion of the reduced
theory by means of restriction.
 Moreover any two $Q$-invariant extrema
$\fti$ and $\fti '$
of $S [ \Phi ]$ give rise to the same extremum of $S_{red} [\f]$
given that their restrictions to $R_{Q}$ coincide,
 $\fti _{R_{Q}} = \fti ' _{R_{Q}}$. 
Note also that $S [\fti ] = S [ \fti ']$ in virtue of 
assumed localization of integrals over $\Sigma$ with
$Q$-invariant integrals. 
Thus we
established a one-two-one correspondence between
instantons of the reduced theory and critical submanifolds
of $E$ consisting of $Q$-invariant instantons of the
original theory having a given restriction to $Q$. 
It follows from above that such BPS-like solutions 
completely determine the instantons contribution
to the $Q$-invariant sector
of the original theory. Really, if we suppose
for example that the instantons $\f _{q}$ 
of the reduced theory are isolated and
classified by an integer $q$, then by virtue of
the equality (15) the instanton contribution to
the partition function of the original (Wick rotated) 
theory  is equal to
\begin{eqnarray}
\sum_{q} \frac{e^{-\beta S_{red} [\f _{q} ]}}
{\sqrt{det Hess (S_{red} [\Phi _{q}])}}, 
\end{eqnarray}
and is clearly determined by $Q$-invariant exrema only.

Our conclusions concerning the dimensional
reduction of supersymmetric field
 theories generalize and provide the 
geometrical understanding of the results of ref. \cite{Cardy}.
 The cited paper contains
the first non-pertubative proof of the dimensional reduction
of Parisi-Sourlas model and describes the relation
between instanton sectors of Parisi-Sourlas model on
a linear $(3, 2)$ space and its reduction which
is a bosonic theory in dimension 1. 

\section{Applications and conclusions.}   

In conclusion let us explain the relation of Parisi-Sourlas
model to the discussion above. 

Consider a supermanifold $\s = B \times {\cal R} ^{(2, 2)}$, 
where $B$ is a (super)manifold. Let $M$ be a
linear superspace. 
Let us choose local coordinates
$\{ x^{i}, y^{\al}, \th , \tb \}$ on $\s$, where
$\{ x^{i} \}$ is a set of local coordinates on $B$, 
$\{ y^{\al}, \th , \tb \}$ are even and odd coordinates
on ${ \cal R } ^{(2, 2)}$. 
Let $h$ be a Riemannian metric on the manifold $B$. 
Then a metric
on $\s$ can be defined by means of the following
quadratic form: 
\begin{eqnarray}
g = h_{ij} (x) \delta x^{i} \delta x^{j} + \sum_{\al}\delta y^{\al}
\delta y^{\al} + 2 \delta \th \delta \tb, 
\end{eqnarray}
Consider the following $\sigma$-model having
$\s$ as a source manifold: 
\begin{eqnarray}
Z [\beta ] = \int [d \Phi ]_{E} e^{i \beta S[\Phi ]}, \\
S[\Phi ] =\int_{\s} dV (g^{-1} 
(\Phi ^{I}, \Phi ^{J}) G_{I~J} (\Phi) +V(\Phi )), 
\end{eqnarray}
 where $E = \{ \s \rightarrow M \}$, $\Phi \in E$; 
$g^{-1}$ is a bivector field on $\s$ inverse
to the quadratic form (25). 
In components $\Phi ^{I} = \phi ^{I} + \psi ^{I} \tb +
\overline{\psi } ^{I} \th + A^{I} \th \tb$. It follows
from the results of  \cite{PS} that the model (26), (27)
can be viewed as a result of stochastic
quantization of a $\sigma$-model defined
on the space of maps 
$\{ B \times {\cal R} ^{(2, 0)}\rightarrow M \}$. 
Corresponding action functional is
 \begin{eqnarray}
S [\phi] = \int_{\sn} dV((h^{ij} 
\frac{\partial \phi ^{I}}{\partial x^{i}} 
\frac{\partial \phi ^{J}}{\partial x^{j}}  + \sum_{\al} 
\frac{\partial \phi ^{I}}{\partial y^{\al}}  
\frac{\partial \phi ^{J}}{\partial y^{\al}}) G_{I~J} (\Phi) 
 + V(\phi)), 
\end{eqnarray}
In such interpretation we consider only the following
correlation functions of the model (26), (27): 
$< \Phi^{I_{1}} \mid _{B} \ldots \Phi^{I_{k}} \mid _{B} >$. 
It is easy to check that the metric form (25) is
invariant with respect to the following odd vector field
on $\s$: 
\begin{eqnarray}
Q= \tb \frac{\partial }{\partial y^{1}} +
 \th \frac{\partial }{\partial y^{2}}
+ y^{2}  \frac{\partial }{\partial \tb} 
- y^{1} \frac{\partial }{\partial \th} 
\end{eqnarray}
This vector field satisfies all 
conditions of  the Corollary and the Theorem 
which yields the integration formula (1) 
for the integrals of $Q$-invariant
functions over $\s$. 

Consider the $Q$-invariant sector of the model (26), (27).  
This sector describes in particular stochastic correlation
functions of the model (28). 
The results of the previous section suggest that
this sector is equivalent to the following model
defined on the manifold $B$: 
\begin{eqnarray}
Z[\beta] = \int [d \phi] _{E_{B}} e^{i S[\phi]} , \\
S[\phi ] = \int_{B} dv (g^{ij} \partial _{i} \phi^{I} 
\partial ^{j} \phi ^{J} G_{I~J} (\phi )
 +V(\phi)), 
\end{eqnarray}
where $E_{B} = \{ B \rightarrow M \}$ and $\phi \in E_{B}$
and $dv$ corresponds to the metric $h$ on $B$. 
This conclusion agrees with corresponding statements about
dimensional reduction of the original Parisi-Sourlas model
and its modifications considered in \cite{Niemi}. 

Finally let us note that all results of the present section can be
generalized to the case when $\s$ is a total space
of a flat $(m, m)$-bundle over the base $B$.

\section{Acknowledgments. }
I am grateful to A. M. Polyakov and A. S. Schwarz
for drawing my attention to the problem and
illuminating discussions afterwards.

\end{document}